\documentclass[osajnl,twocolumn,showpacs,superscriptaddress,10pt]{revtex4-1} 
\usepackage{amsmath,amssymb,graphicx}
\begin{document}

\title{Dynamical generation of dark solitons in spin-orbit-coupled Bose-Einstein condensates }

\author{Shuai Cao}
\affiliation{College of Sciences, South China Agricultural
University, Guangzhou, 510642, China} \affiliation{Laboratory of
Quantum Engineering and Quantum Materials, and School of Physics and
Telecommunication Engineering, South China Normal University,
Guangzhou 510006, China}

\author{Chuan-Jia Shan}
 \affiliation{Laboratory of Quantum
Engineering and Quantum Materials, and School of Physics and
Telecommunication Engineering, South China Normal University,
Guangzhou 510006, China}
\affiliation{College of Physics and Electronic Science, Hubei Normal
University, Huangshi 435002, China}

\author{Dan-Wei Zhang}
\email{Corresponding author: zdanwei@126.com} \affiliation{Laboratory of Quantum
Engineering and Quantum Materials, and School of Physics and
Telecommunication Engineering, South China Normal University,
Guangzhou 510006, China}
\affiliation{Department of Physics and
Center of Theoretical and Computational Physics, The University of
Hong Kong, Pokfulam Road, Hong Kong, China}

\author{Xizhou Qin}
\affiliation{State Key Laboratory of Optoelectronic Materials and
Technologies, School of Physics and Engineering, Sun Yat-Sen
University, Guangzhou 510275, China }

\author{Jun Xu}
\affiliation{State Key Laboratory of Optoelectronic Materials and
Technologies, School of Physics and Engineering, Sun Yat-Sen
University, Guangzhou 510275, China }

\begin{abstract}We numerically investigate the ground state, the Raman-driving
dynamics and the nonlinear excitations of a realized
spin-orbit-coupled Bose-Einstein condensate in a one-dimensional
harmonic trap. Depending on the Raman coupling and the interatomic
interactions, three ground-state phases are identified: stripe,
plane wave and zero-momentum phases. A narrow parameter regime with
coexistence of stripe and zero-momentum or plane wave phases in real
space is found. Several sweep progresses across different phases by
driving the Raman coupling linearly in time is simulated and the
non-equilibrium dynamics of the system in these sweeps are studied.
We find kinds of nonlinear excitations, with the particular dark
solitons excited in the sweep from the stripe phase to the plane
wave or zero-momentum phase within the trap. Moreover, the number
and the stability of the dark solitons can be controlled in the
driving, which provide a direct and easy way to generate dark
solitons and study their dynamics and interaction properties.
\end{abstract}


\maketitle 

\section{Introduction}

Spin-orbit (SO) coupling plays an important role in many fundamental
quantum phenomena, ranging from the atomic fine structure to the
newly discovered topological insulators \cite{Hasan,Qi}. Recently, a
synthetic SO coupling has been successfully engineered in
Bose-Einstein condensations (BECs) and degenerate Fermi gases
\cite{Lin2011,Williams,Zhang2012,Qu,LeBlanc,Olson,Zhang2014,Wang1,Cheuk,Wang2}.
These cold atomic systems with adjustable SO coupling provide an
ideal platform for investigating a wide range of interesting physics
in many SO-coupled systems, such as atomic spin Hall effects
\cite{Zhu2006,Beeler}, Anderson localization of relativistic
particles \cite{Zhu2009}, fractional Fermi number \cite{Zhu2012} and
topological superfluid with Majorana fermions \cite{Zhu2011}. The
cold atoms with synthetic SO-coupling also have potential
applications in designing atomic interferometry \cite{Anderson2011}.

Since the newly realized SO-coupled BEC has no direct analog in
solids, great attention has been paid to this unique many-body
system. Especially, the ground state of SO-coupled BECs has been
studied extensively in theory \cite{Wang, Ho, Wu, Li, Jian, Hu, Xu,
Ozawa, Sinha, Zhou, Radic, Xu2, Yu, Ramachandhran, He,Ozawa2,
Zhou2,Lewenstein,Zhu2013,Zhao2014}. For a two-dimensional
homogeneous bosonic gas with Rashba SO coupling, depending on the
interatomic interactions, the ground state of the system exhibits
two unconventional phases: a plane wave phase with condensate in a
single momentum state and a stripe phase with condensate in two
opposite momenta \cite{Wang, Ho, Wu}. While for the realized
SO-coupled BEC with equal contributions of Rashba and Dresselhaus SO couplings
\cite{Lin2011,Williams,Zhang2012,Qu,LeBlanc,Olson,Zhang2014}, it is
found that the ground state belong to the two phases (the stripe
phase and the plane wave phase) or the conventional zero-momentum
phase, and thus an interesting tri-critical point was predicted in
the phase diagram \cite{Li}. Taking the finite temperature
\cite{Jian}, a harmonic trap \cite{Hu,Xu,Ozawa} or a rotating trap
\cite{Sinha, Zhou, Radic, Xu2} into account, more exotic phases were
revealed, such as half vortices and Skyrmion lattices. Loading the
SO-coupled bosons into optical lattices, many novel magnetic ground
states were also found \cite{Lewenstein,Zhu2013,Zhao2014}.

The dynamics of SO-coupled BECs has also been studied in different
contexts \cite{Zhang2012,Duine,Tokatly,Natu,Josephson,Zhang,KT}. For
example, they have been demonstrated to exhibit unconventional
collective dipole oscillations \cite{Zhang2012,Duine} and spin
dynamics \cite{Tokatly, Natu}, interesting spin Josephson effects
\cite{Josephson} and relativistic dynamics with analogs of
Zitterbewegung \cite{Zhang} and Klein tunneling \cite{KT} under
certain conditions. For the realized Rashba-Dresselhaus SO-coupled
BEC, if the Raman coupling is driven in time, the system will pass
through phase transition points. However, such Raman-driving
dynamics in this system is yet to be explored.

On the other hand, the interatomic interactions in BECs lead to
inherent nonlinearity under the Gross-Pitaevskii mean-field
description. Thus the dynamics of a BEC is governed by a nonlinear
Schr\"{o}dinger equation, where the interaction strength could be
tuned by Feshbach resonances \cite{Chin}. With the tunable
nonlinearity, the dark solitons \cite{Fialko,Achilleos1}, bright
solitons \cite{Achilleos2, yongXu}, and gap solitons
\cite{Kartashov} in SO-coupled BECs are also investigated recently.
Interestingly, the dark solitons can be excited by other methods.
One can engineer a phase difference in a condensate to create a dark
soliton at the interface between the phase domains \cite{Burger} or
merge two coherent condensates to create multiple dark solitons
\cite{Weller}. Other schemes of generating dark solitons involve
driving the system away from equilibrium \cite{Zurek, Damski}, such
as using nonlinearity-assisted quantum tunneling \cite{Lee}. Thus, a
natural question is whether we can dynamically generate the solitons
in the realized SO-coupled BEC by the Raman-driving.

In this paper, we numerically investigate the Raman-driving dynamics
and the nonlinear excitations of the realized SO-coupled BEC in a
one-dimensional (1D) harmonic trap. By using imaginary time
evolution method, we first obtain the ground state of the system and
identify the three quantum phases depending on the Raman coupling
strength and the interatomic interactions. Within the external trapping
potential, we find a narrow parameter regime with coexistence of
stripe and zero-momentum or plane wave phases in real space, which is absent in the homogenous case \cite{Li}.
With the operator-splitting procedure, we then simulate four different sweeps
across the ground-state phases by driving the Raman coupling
linearly in time. In these sweeps, we study the non-equilibrium
dynamics of the system and obtain different nonlinear excitations.
Interestingly, we find that dark solitons can be excited within the
trap in the sweep from the stripe phase to the plane wave or
zero-momentum phase. Moreover, we show that the number and the
stability of the dark solitons can be controlled in the dynamical
progress, providing a direct and easy way to create dark solitons
and study their dynamics and interaction properties.

The remaining part of this paper is organized as follows. We
introduce the model for a SO-coupled BEC in a harmonic trap in Sec.
II. The ground-state properties of the system are numerically
studied in Sec. III. Then we investigate the Raman-driving dynamics
and excitations in the sweeps across different ground-state phases
in Sec. IV. Finally, a short summary is given in Sec. V.

\section{The model}

We consider a SO-coupled BEC confined in a quasi-1D harmonic trap
with equal strength of Rashba and Dresselhaus SO couplings,
which has been realized with $^{87}$Rb atoms
\cite{Lin2011,Williams,Zhang2012,Qu,LeBlanc,Olson,Zhang2014}. In
the mean-field approach, the energy functional of the system is $E =
\int_{-\infty}^{+\infty} \mathcal{E}dx$, with the energy density
\begin{equation}
 \mathcal{E} = \frac{1}{2}\left(\Psi^{\dag}\mathcal{H}_0\Psi+g_{\uparrow\uparrow}|\psi_{\uparrow}|^4+g_{\downarrow\downarrow}|\psi_{\downarrow}|^4
 +2g_{\uparrow\downarrow}|\psi_{\uparrow}|^2|\psi_{\downarrow}|^2\right),
\label{h0}
\end{equation}
where $\Psi\equiv(\psi_{\uparrow},\psi_{\downarrow})^T$ with
$\psi_{\uparrow}$ and $\psi_{\downarrow}$ being the two (pseudo-)
spin wave functions of the BEC, and the effective 1D interaction
constants $g_{\sigma\sigma'}=2\hbar^2
a_{\sigma\sigma'}N/(ml_{\perp}^{2})$ are defined by the s-wave
scattering lengths $a_{\sigma\sigma'}$ with
$\sigma,\sigma'=\uparrow,\downarrow$, the particle number $N$ and
the oscillator length associated with a harmonic vertical
confinement $l_{\perp}$. Hereafter we assume
$g_{\uparrow\uparrow}=g_{\downarrow\downarrow}\equiv g_0$ and
$g_{\uparrow\downarrow}=g_{\downarrow\uparrow}\equiv g_1$ for
simplicity. The single particle Hamiltonian $\mathcal{H}_0$ in Eq.
(1) is given by \cite{Lin2011}
\begin{equation}
\mathcal{H}_0 = \frac{1}{2m}\left[\left(\hat{p}_x-k_0
\sigma_z\right)^2 \right] + \frac{\Omega}{2} \sigma_x +
\frac{\delta}{2} \sigma_z +V_{\text{ext}}, \label{h0}
\end{equation}
where $m\simeq1.44\times10^{-25}$ kg is $^{87}$Rb atomic mass,
$\hat{p}_x=-i\hbar\partial_x$ is the momentum operator, $k_0$ is the
wave number of the Raman lasers for coupling the two hyperfine
states, $\sigma_{x,z}$ are Pauli matrices, $\Omega$ is the Raman
coupling strength, $\delta$ is an effective Zeeman field for the
spin states, and $V_{\text{ext}}= m\omega_x^{2}x^2/2$ is the
external trapping potential with frequency $\omega_x$. We
assume the typical trapping frequency $\omega_x=2\pi\times20$ Hz,
such that $a_x=2.4$ $\mu$m. We further assume $l_{\perp}=0.1a_x$,
$k_0=5a_x^{-1}$, and the total number of atoms $N\approx10^5$. Then
the interaction strengths $g_0$ and $g_1$ are on the range from several tens to several hundreds of $\hbar\omega_x$ for
typical scattering lengths.

Measuring the length in units of $a_x=\sqrt{\hbar/(m\omega_{x})}$, time in units
of $\omega_{x}^{-1}$, energy in units of $\hbar\omega_{x}$, we
derive the dimensionless equations of motion for wave functions
$\psi_{\uparrow,\downarrow}$ \cite{Achilleos1}:
\begin{eqnarray}
 i\partial_t \psi_{\uparrow}&=&\frac{1}{2}(-i\partial_x-k_0)^{2}\psi_{\uparrow}+\frac{\delta}{2}\psi_{\uparrow} +
 V_{\text{ext}}\psi_{\uparrow}+ \nonumber\\ &&
 (g_0| \psi_{\uparrow} |^2+ g_1| \psi_{\downarrow} |^2 ) \psi_{\uparrow} +\frac{\Omega
 }{2}\psi_{\downarrow},\\
 i\partial_t \psi_{\downarrow}&=&\frac{1}{2}(-i\partial_x+k_0)^{2}\psi_{\downarrow}-\frac{\delta}{2}\psi_{\downarrow}+
  V_{\text{ext}}\psi_{\downarrow}+ \nonumber\\ &&
  (g_0| \psi_{\downarrow} |^2+ g_1| \psi_{\uparrow} |^2) \psi_{\downarrow}+\frac{\Omega }{2}\psi_{\uparrow}.
\end{eqnarray}
Here we have used $\omega_x\rightarrow\omega_x/\omega_x=1$, $k_0\rightarrow a_xk_0$, $\delta\rightarrow \delta/(\hbar\omega_x)$ and $\Omega\rightarrow \Omega/(\hbar\omega_x)$, thus now $V_{\text{ext}}=\frac{1}{2}x^2$.

In the following, we numerically minimize the energy functional to
obtain the ground-state wave functions of the system described by
Eq. (1) and Eqs. (3,4) for varying Raman coupling strength using the
imaginary time evolution method. Then in the next section, we
simulate the Raman-driving dynamics across different ground-state
phases by integrating the equations of motion with the
well-developed operator-splitting procedure. In the following numerical
calculations, we will mainly consider $\delta=0$ for simplicity.

In simulations, we fix $k_0=5$ and mainly consider
two typical groups of interaction parameters: $g_0=100$ and
$g_{1}=80$ as the first case, $g_0=500$ and $g_{1}=100$ as the
second case. The reason for our choice is that they respectively
give rise to the three and two ground-state phases and meanwhile
within the typical experimental parameter range. However, we note that the
ground-state properties and the Raman-driving dynamics with excited
dark solitons, which will be discussed in the following two
sections, mostly remain for different choices of interaction
parameters. In experiments, the values of $g_0$
and $g_1$ may be directly changed by varying the transverse trapping
strength, but the ratio $g_0/g_1$ is fixed in this way.

\begin{figure}[tbp]
\includegraphics[width=6cm]{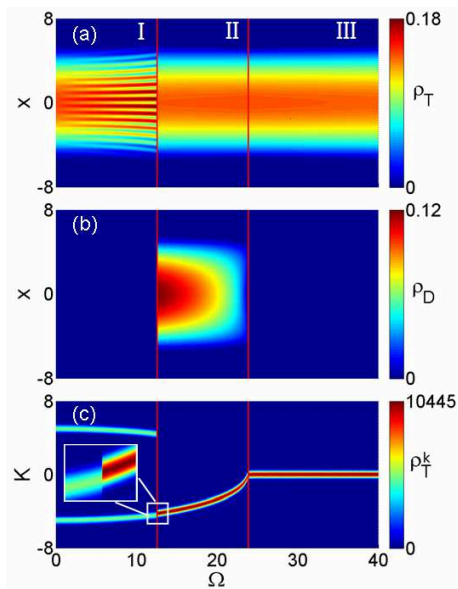}
\caption{(Color online) The ground states of the SO-coupled BEC for
different Raman coupling strength $\Omega$ with the dimensionless
parameters: $k_0=5$, $g_0=100$ and $g_1=80$. (a) The
total density distribution of the ground state $\rho_T(x)$ in real
space. (b) The difference density distribution of the two spin
components $\rho_D(x)$ in real space. (c) The momentum distribution
of the total density for the case in (a). The horizontal red lines
correspond to the boundaries between different phases. Phases I, II,
and III respectively denote the stripe phase, the plane wave phase,
and the zero-momentum phase.
The characteristics of the ground states in the three phases are in the text.  
}
\end{figure}

\begin{figure}[tbp]
\includegraphics[width=6cm]{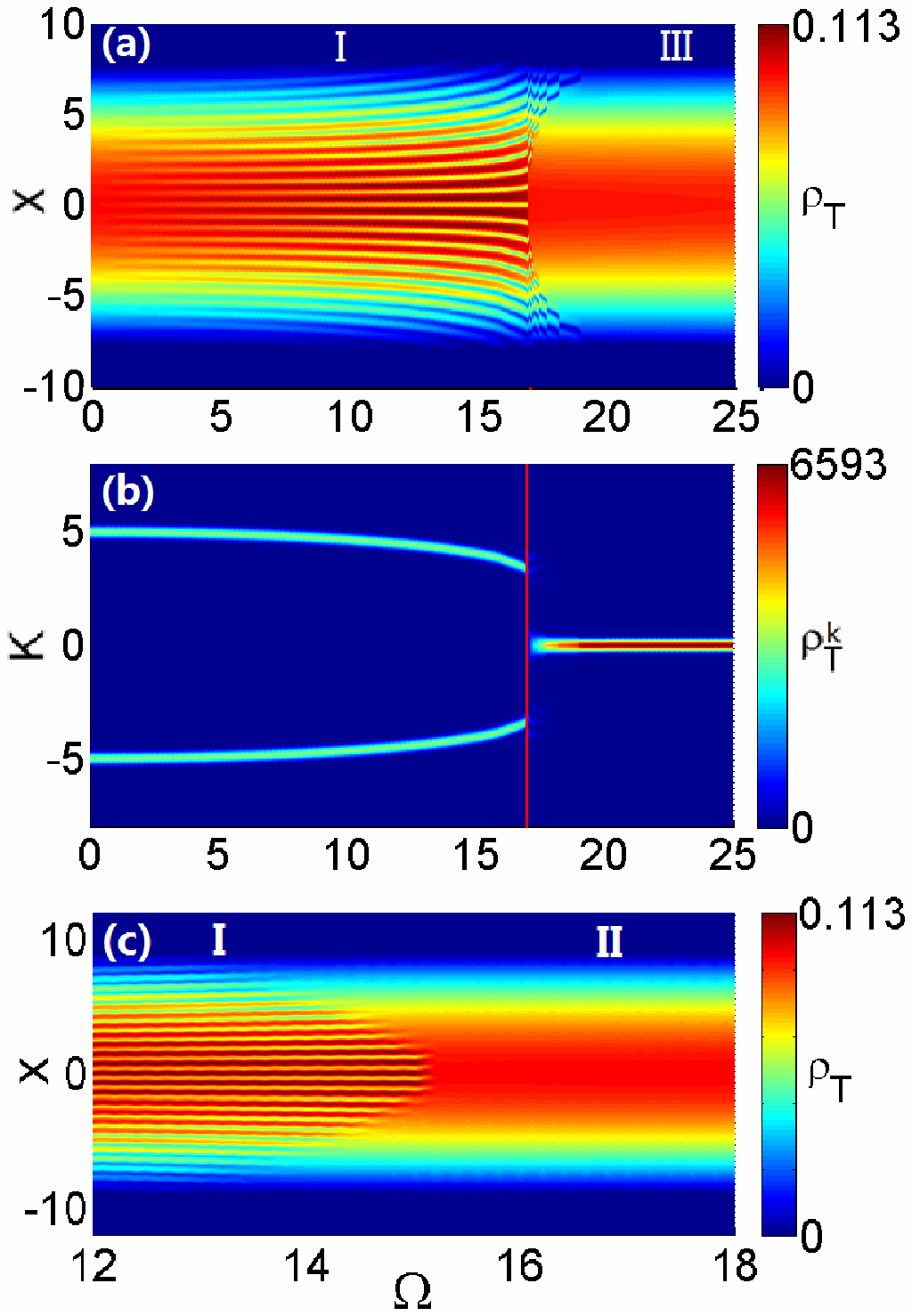}
\caption{(Color online) The ground states of the SO-coupled BEC for
different Raman coupling strength $\Omega$ with the dimensionless
interaction parameters: (a,b) $g_0=500$ and $g_1=100$; (c) $g_0=500$
and $g_1=400$. (a,c) The total density distribution of the ground
state $\rho_T(x)$. (b) The corresponding momentum distribution of
the total density. The horizontal red lines correspond to the
boundary between phases I and III. The coexistence of phases I and
III in a small region near the critical Raman coupling strength is
characterized by Gaussian density distribution in the trap center
with stripe density at the edge as shown in (a), and the coexistence
of phases I and II is characterized by Gaussian density distribution
at the edge with stripe density in the center as shown in (c). Other
parameters in (a-c) is $k_0=5$.}
\end{figure}

\section{Ground states}

The system in the uniform case ($V_{\text{ext}}=0$) has been well
studied in Ref. \cite{Li}, and it is analytically found that if the
SO-coupling dominates and the condition $k_0^2>G_-+\frac{G_-}{4G_+}$
is satisfied with $G_{\pm}\equiv g_{0}\pm g_{1}$, there are three
different phases with interesting condensate states for varying
Raman coupling strength \cite{Li}. However, if the atomic
interactions dominate and the condition is not satisfied, there are
two condensate states \cite{Li}. Even though the boundary condition
can not be solved analytically in the presence of a trapping
potential, our simulations demonstrate that the two cases also
appear. The numerical results of the ground states are shown in
Fig. 1 and Fig. 2, corresponding to the two groups of interaction
parameters, respectively. Here we plot the total density for the
ground state wave functions $\rho_T$ and the difference density
between two spin components $\rho_D$ (spin polarization) in real space as a function of
Raman coupling strength, with
\begin{eqnarray}
\rho_T \equiv| \psi_{\uparrow}(x)|^2 +| \psi_{\downarrow}(x) |^2,\nonumber \\
\rho_D\equiv|\psi_{\uparrow}(x)|^2 -| \psi_{\downarrow}(x) |^2.
\end{eqnarray}
We also plot the total density in the momentum space
$\rho_T^{k}\equiv| \psi_{\uparrow}(k)|^2 +| \psi_{\downarrow}(k)
|^2$, where the wave function
$\psi_{\sigma}(k)=\frac{1}{\sqrt{2\pi}}\int_{-\infty}^{+\infty}\psi_{\sigma}(x)e^{ikx}dx$.
We will see that $\rho_T$, $\rho_D$ and $\rho_T^k$ together
characterize different condensate states, with the central momentum
of the condensate and the spin polarization being the order
parameters \cite{Li}.

\begin{figure}[tbp]
\includegraphics[width=8.8cm,]{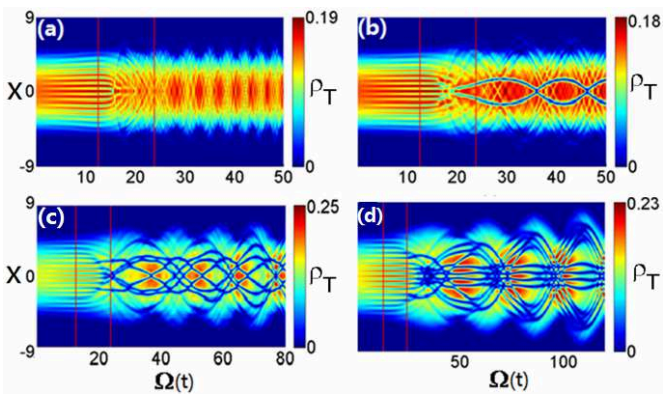}
\caption{(Color online) The density evolution $\rho_T(x,t)$ in the sweep from phase I to phase II by Raman driving
$\Omega(t)=\Omega_0+\beta t$ with $\Omega_0=0.5$ and typical sweep
rates $\beta$: (a) $\beta=0.6$, (b) $\beta=1.5$, (c) $\beta=3$ and
(d) $\beta=6$. The horizontal red lines corresponds to the phase transition points. Other parameters
are $k_0=5$, $g_0=100$ and $g_1=80$. }
\end{figure}

We first consider the case that the SO-coupling dominates with
$k_0=5$, $g_0=100$ and $g_{1}=80$. As shown in Fig. 1, three
ground-state phases are respectively denoted by phases I, II and III
with increasing $\Omega$, and their boundaries are marked by the red
solid lines. Phase I is the stripe phase for relatively small
$\Omega$, characterized by fringes in the density $\rho_T$ in Fig.
1(a). In this phase, the atoms condense in an equal-probability
superposition of two plane waves with opposite wave vectors as shown
in Fig. 1(c). Besides, the spin polarization identically vanishes,
leading to $\rho_D=0$ in Fig. 1(b). It is interesting to note that if
$\delta\neq0$, the spin symmetry will be broken and then the phase separation
of the ground-state BEC can appear in the regime of phase I, which has been observed in experiments \cite{Lin2011,Zhang2012}.
We have numerically confirmed the appearance of phase separation for finite $\delta$ and above a critical Raman coupling strength in the regime of Phase I, which is not shown in Fig. 1 for the $\delta=0$ case.

Phase II is the plane wave phase for intermediate $\Omega$, characterized by a single non-zero
momentum peak in $\rho_{T}^k$ and finite spin polarization. Phase
III is the zero-momentum phase for relatively large $\Omega$, where
the central momentum of the condensate is zero and the spin
polarization also vanishes. The phase transition between I and II
belongs to the first order phase transition with a jump in the
central momentum as shown in the insert figure in Fig. 1(c). While
the one between II and III belongs to the second order phase
transition. In all of the three phases, the overall density
configuration exhibits a Gaussian distribution due to the external
harmonic trap, which also demonstrates that the local density
approximation works well here.

For the case that the interactions dominate with $k_0=5$, $g_0=500$
and $g_{1}=100$, the results of the ground states are shown in Fig.
2. In this case, we can see that there are only phases I and III.
The phase transition between I and III belongs to the first order
phase transition. Interestingly, we find the coexistence of the two
phases in a small region near the critical Raman coupling strength,
characterized by Gaussian density distribution in the trap center
with stripe density in the edge as shown in Fig. 2(a). This
phenomenon is due to the inhomogeneous interaction effect by the
trapping potential, which is absent in the homogenous case \cite{Li}.
In the local density approximation, the density distribution of
atoms is inhomogeneous and decrease away from the center of the
trap, leading to the decreasing interaction strength from the center
to the edge. Since the interactions play an crucial role in this
case and the critical Raman coupling strength between phases I and
III depends on the interaction strength, the central part of the
condensate will enter into the phase III earlier than the edge when
increasing Raman coupling.

One may also expect a coexistence of stripe and plane wave phases in
real space from the same mechanism, which is absent for the case in
Fig. 1 due to the weak interactions. Actually, for appropriately
intermediate interaction strengths, the I-II coexistence phase is
indeed found in our numerical simulations. An example is shown in
Fig. 2(c) for $g_0=500$ and $g_1=400$, where the I-II coexistence
phase is characterized by Gaussian density distribution at the edge
with stripe density in the center. This result further demonstrates that the
two coexistence phases are not always present in the whole interaction
parameter regime and actually from the competitions between repulsive
interactions and external trapping.

\begin{figure}[tbp]
\includegraphics[width=8.8cm,]{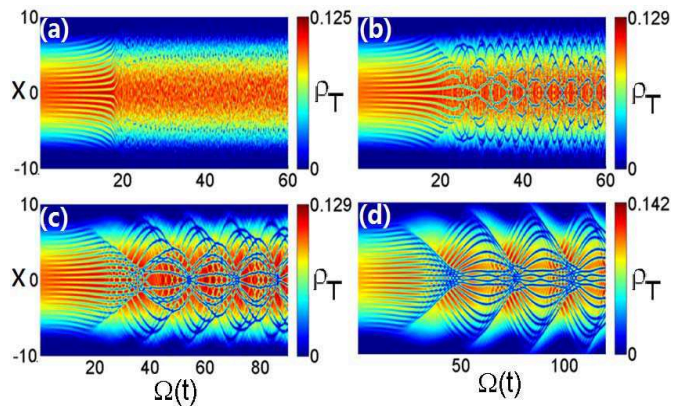}
\caption{(Color online) The density evolution $\rho_T(x,t)$ in the sweep from phase I to phase III by Raman
driving $\Omega(t)=\Omega_0+\beta t$ with $\Omega_0=0.5$ and typical
sweep rates $\beta$: (a) $\beta=0.1$, (b) $\beta=0.6$, (c) $\beta=3$ and (d) $\beta=6$.
Other parameters are $k_0=5$, $g_0=500$, and $g_1=100$. }.
\end{figure}

\section{Raman-driven dynamics and generation of dark solitons}

In this section, we consider the quantum dynamics and excitations of
the SO-coupled BEC in the sweeps between three ground-state phases.
To do this, the system is initially prepared in one of its ground
states, and is then driven across the phase transition points by the
Raman-driving, which can be achieved by changing the intensity of
the Raman lasers in experiments. We consider that the Raman coupling
strength varies linearly in time $t$:
\begin{eqnarray}
 \Omega(t)= \Omega_0+\beta t,
\end{eqnarray}
where $\Omega_0$ is the Raman coupling strength at $t=0$ and $\beta$
is the Raman-driving rate. The ground states depend on
$\Omega_0$, $g_{0}$ and $g_1$, with the phase boundaries shown in
Figs. 1 and 2. In the following, we numerically simulate these
sweeping progresses by integrating the equations of motion (3) and
(4).

\subsection{Dark solitons in I-II and I-III sweeps}

We first consider the sweep from phase I to phase II with typical
interaction strengths $g_0=100$ and $g_1=80$. At the outset the
Raman coupling $\Omega_0=0.5$, and the atoms condense into the
stripe ground state within the region of phase I, which is obtained
numerically by the imaginary time evolution method. Then the Raman
coupling strength is increased as the form given by Eq. (6). The
numerical results of the Raman-driving dynamics for some typical
sweeping rates $\beta$ are shown in Fig. 3, where we plot the
density evolution of the SO-coupled BEC $\rho_T(x,t)$. Here the two horizontal red lines denote the phase
boundaries. For slow sweep with $\beta=0.6$ as shown in Fig. 3(a),
the BEC begins to oscillate after passing through the first phase
transition point between I and II and then gradually exhibits
breathing oscillation. The amplitude and frequency of the breathing
oscillation characterize the propagation of excitation modes, which
are different from the sound modes in this system when subjected to
a perturbation driving \cite{Li}. We have also checked that in the
very slow sweeping limit, the system will evolve adiabatically without
collective excitations (which is not shown here).

When the sweep is fast enough with $\beta=1.5$, we find that a pair
of dark solitons exhibit in the BEC after passing through the phase
transition point as shown in Fig. 3(b). The dark solitons are generated
in the center because the nonlinear interactions are stronger there within
the external harmonic trap. Due to the harmonic force, the solitons are reflected from the edges
of the trap and collide in the center again and again, giving rise
to the oscillation behavior of the dark solitons and interference
pattern in Fig. 3(b). When increasing the sweep rate with
$\beta=3,6$ as shown in Figs. 3(c,d), more and more dark soltions
are excited in the system with similar oscillation dynamics and
interference pattern.

We then consider the sweep from phase I to phase III with typical
interaction strengths $g_0=500$ and $g_1=100$, which is also the
first order phase transition. We also set the system initially in
the ground state with $\Omega_0=0.5$ and the Raman coupling is
increased linearly in time. The numerical results of the evolution
of atomic density $\rho_{T}(x,t)$ are shown in Fig. 4 for
different sweep rates. For very slow sweep with $\beta=0.1$ as shown
in Fig. 4(a), neither dark solitons nor breathing modes are excited
in the BEC in this case. While for $\beta=0.6$ in Fig. 4(b), we find
that more than one pair of dark solitons are generated compared to
the I-II sweep with the same sweeping rate in Fig. 3(a). This is due
to the facts that the stronger atomic interaction in this case leads
to more density dips for the dynamical generation of dark solitons.
In this case, also more and more dark solitons are excited with
similar oscillation and interference dynamics when increasing the
sweep rate with $\beta=3,6$ as shown in Figs. 4(c,d). In a word,
the dark solitons are more easily excited in the I-III sweep than in
the I-II sweep.

We should note that the boundaries among the three ground-state
phases and the corresponding spin polarization will be modified for the case $\delta\neq0$, which
corresponds to broken spin symmetry \cite{Li}. However, we have
numerically confirmed that the dark solitons can also be created in
these non-adiabatic sweeps for $\delta\neq0$. Moreover, this way of
generation of dark solitons is not specific to the particular
choices of interaction and SO-coupling strengths as those in Figs. 3
and 4. As long as the sweeps are from the stripe ground state and
fast enough, the dark solitons will be excited and exhibit
oscillatory motion and collision in the trap. In contract, the
existence of solitons in SO-coupled BECs in previous work closely
relies on certain particular conditions, such as ring-shape trapping
potentials \cite{Fialko}, stationary points in the dispersion
relation \cite{Achilleos1}, attractive interactions
\cite{Achilleos2, yongXu} and spatially periodic Zeeman fields
\cite{Kartashov}. Therefore, our work demonstrated a general method
of dynamically generating solitons in SO-coupled BECs, which is easy
to be implemented in current experiments without carefully designing
the Raman lasers and atomic interactions.

\begin{figure}[tbp]
\includegraphics[width=8cm,]{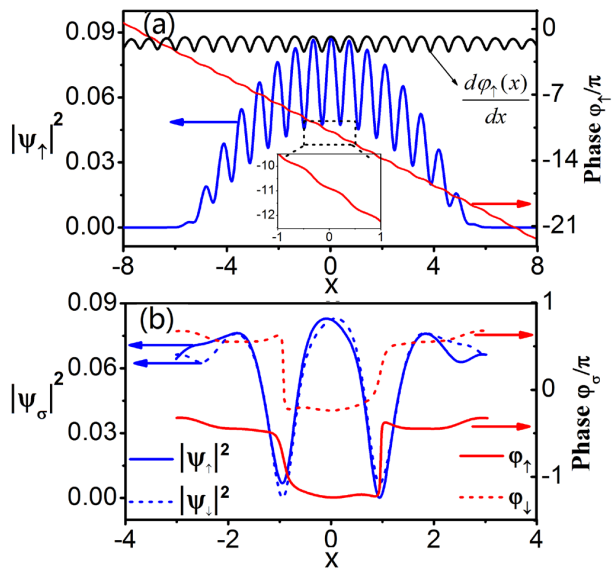}
\caption{(Color online) (a) The spin-up-component density
$|\psi_{\uparrow}(x)|^2$ (blue solid) and phase
$\varphi_{\uparrow}(x)$ (red solid) of the stripe ground state for
$\Omega=12$. The black solid line denotes the spatial derivative of
the phase $d\varphi_{\uparrow}(x)/dx$, which shows maximum phase
variation at the density dips. (b) The density
$|\psi_{\sigma}(x)|^2$ (blue solid line for $\sigma=\uparrow$ and
blue dashed line for $\sigma=\downarrow$) and phase
$\varphi_{\sigma}(x)$ (red solid line for $\sigma=\uparrow$ and red
dashed line for $\sigma=\downarrow$) of the excited dark solitons
when $\Omega(t)=50$ in Fig. 3(b). Other parameters are
$k_0=5$, $g_0=100$, $g_1=80$ and $\beta=1.5$.}
\end{figure}

\subsection{Formation, stability and dynamics of the excited dark solitons}

In this part, we further discuss the formation, stability and
dynamics of the excited dark solitons in the sweeps. In experiments,
one could use the phase imprinting and density engineering methods to
create dark solitons in an ordinary BEC \cite{Burger,Weller}.
However, the different mechanism of dynamically generating dark
solitons here is by sweeping the BECs away from equilibrium stripe
ground states and across phase boundaries \cite{Zurek, Damski}. In a
harmonically trapped condensate with repulsive interatomic
interactions, the nodes of the excited nonlinear eigenstates can
evolve into the dark solitons \cite{Lee,Kivshar}.

For the stripe condensate, there are many dips in its density
distribution to support the formation of dark solitons when the
system is driven fast enough from phase I to phase II, as shown in
Figs. 3(b-d). To see this point more clearly, we first plot the
spin-up-component density $|\psi_{\uparrow}(x)|^2$ and the phase
$\varphi_{\uparrow}(x)$ of the stripe ground state in Fig. 5(a).
Even though the phase is relatively smooth in space, the maximum
phase variation $d\varphi_{\uparrow}(x)/dx$ locates at the density
dips, which is favorable to form dark solitons. The situations of
the spin-down component are similar. We then plot the spin density
$|\psi_{\sigma}(x)|^2$ and the phase $\varphi_{\sigma}(x)$ of the
excited dark solitons [when $\Omega(t)=50$ in Fig. 3(b)] in Fig.
5(b), where the density dips and the accompanied phase jumps of the
two dark solitons can be seen clearly.

\begin{figure}[tbp]
\includegraphics[width=7cm,]{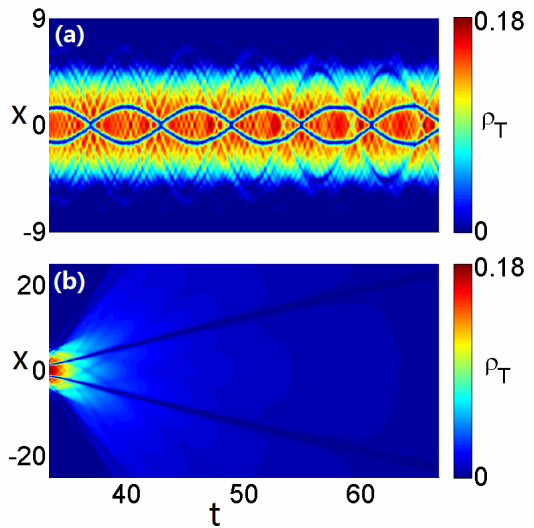}
\caption{(Color online) The stability of the dark solitions in Fig.
3(b). The long-time evolution of the total density $\rho_{T}(x,t)$
for: (a) keeping on the trapping potential; and (b) turning off the
trapping potential, after stopping the Raman driving at time
$t_d=33$. Other parameters are $k_0=5$, $g_0=100$,
$g_1=80$, $\Omega_0=0.5$ and $\beta=1.5$.}
\end{figure}

\begin{figure}[tbp]
\includegraphics[width=8.5cm,]{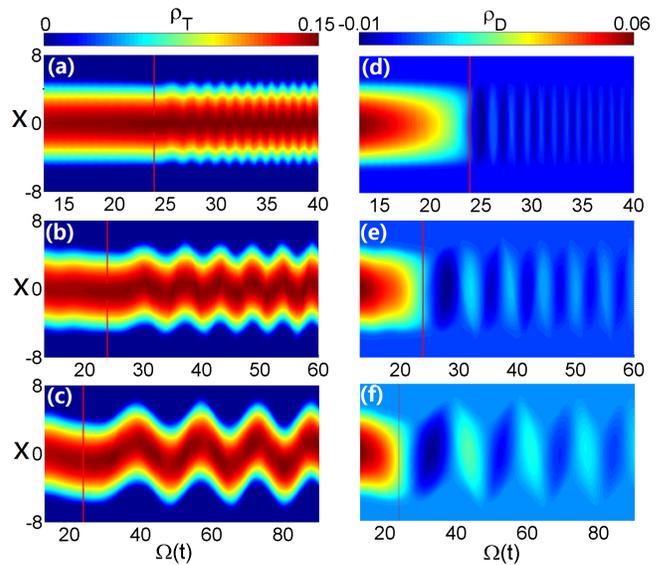}
\caption{(Color online) The density evolutions $\rho_T(x,t)$ and
$\rho_D(x,t)$ in the sweep from phase II to phase III by Raman
driving $\Omega(t)=\Omega_0+\beta t$ with $\Omega_0=13$ and typical
sweep rates $\beta$: (a) and (d) $\beta=0.1$, (b) and (e)
$\beta=0.6$, (c) and (f) $\beta=2$. The horizontal red lines
correspond to the phase transition point. Other parameters are $k_0=5$, $g_0=100$ and $g_1=80$. }
\end{figure}

To verify the stability of the generated dark solitons in Fig. 6, we
calculate the long-time evolution of atomic density $\rho_T(x,t)$
after the driving time $t_d=33$, corresponding to $\Omega(t_d)=50$
and the same parameters in Fig. 3(b). Figures 6(a) and 6(b)
correspond to the cases of keeping on and tuning off the trap when
the driving is stopped at $t=t_d$, respectively. It is clear from
Fig. 6(a) that the dark solitons remain periodically oscillating in
the trap after stopping driving, without significant decay for a
long time. Therefore we can conclude that the dark solitons in this
system are stable within the trapping potential. If the trap is also
turned off with the driving, the dark solitons will propagate and
spread out, which is shown in Fig. 6(b). 

For an equilibrium SO-coupled BEC without driving, the analytical
single-dark-soliton solution of Eqs. (3,4) can be approximately
obtained as the forms in Ref. \cite{Achilleos1}. However, for the
non-equilibrium system with multiple interacting dark solitons here, it is
too complicated to get the analytical forms and derive an effective
model to exactly describe their dynamics as shown in Figs. 3 and Fig.
4. In Ref. \cite{Achilleos1}, the size of a single dark soliton is characterized to be proportional to an effective healing length, which is determined by the atomic interaction strengths and the stationary points in the dispersion relation. In numerical simulations for our case, we can not find a clear and direct connection between the size of excited multiple dark solitons and the (effective) healing length. For the simplest two-soliton case in Figs. 3(b) and 6(a), the
dark solitons clearly show particle-like behaviors with oscillatory
motion and mutually interactions within the external trap, which has
been observed in ordinary BECs \cite{Weller}. For well-separated
dark solitons, their center $X_0(t)$ approximately satisfies the
equation of motion in the form $d^2 X_0/dt^2=-\frac{1}{2}M_0\omega_s^2X_0$,
where $M_0$ is the effective mass of the dark solitons \cite{Achilleos1} and
$\omega_s$ is the oscillation frequency determined by the harmonic
trap \cite{Burger,Weller}. Due to the interactions between dark
solitons, which can be treated by additional potentials, the oscillation frequencies of interacting are modified \cite{Weller}.
In addition, as the non-adiabatic driving does, the motion and collision of multiple dark solitons gradually
generate excitation modes in the SO-coupled BEC, resulting in the
complicated breathing-like behavior in the background as shown in
Figs. 3(c,d) and 4(c,d). Clearly, the oscillation amplitude of the
multiple dark solitons does not remain constant in this case.

\subsection{Dynamics in sweeps between phases II and III}

Now we turn to consider the sweeps between phase II and phase III.
The first case we considered is the sweep from phases II to III,
which belongs to the second order phase transition. To do this, we
set the system initially in phase II with $\Omega_0=13$ and choose
the same interaction parameters as those in Fig. 3: $g_0=100$ and
$g_1=80$. Then the Raman coupling is also increased linearly in
time. In this case, the collective center-of-mass (COM) motion in
$\rho_T(x,t)$ and the periodical oscillation of the spin polarization in
$\rho_D(x,t)$ can emerge, with two examples being shown in Figs.
7(a,d) and 7(b,e) for $\beta=0.1$ and $\beta=0.6$, respectively. It
is clear to see from Fig. 7 that for very slow sweep ($\beta=0.1$),
the system almost follows its ground state in the beginning and then
exhibits the dynamical oscillation once passing through the phase
transition point. In addition, the period and the amplitude of the
oscillation are both increased by increasing the Raman driving rate.
However, we find that the dark solitons can not be excited by the
Raman-driving even in the fast sweep, such as $\beta=2$ in Figs.
7(c,f), which just exhibit similar oscillating COM motion and spin
dynamics. We have also checked that no dark soliton can be excited
in a much faster sweep in this case. This is due to the absence of
density dips in the initial plane-wave state to evolve into
dark-soliton excitations. Since the dark solitons can be excited only from the stripe state, the appearance of dark solitons in the Raman driving is an indication of the existence of stripe state, which has not directly been imaged in experiments \cite{Lin2011,Zhang2012}.

\begin{figure}[tbp]
\includegraphics[width=8.5cm,]{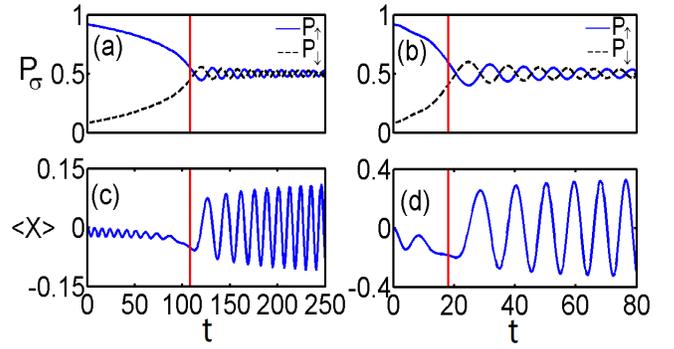}
\caption{(Color online) The population evolution of the spin
components $P_{\sigma}(t)$ in (a) and (b), and the center-of-mass
evolution in (c) and (d) for the II-III sweep. The horizontal red
line corresponds to the phase transition point. The driving rates
are $\beta=0.1$ in (a) and (c), and $\beta=0.6$ (b) and (d). Other
parameters in (a-d) are $\Omega_0=13$, $k_0=5$,
$g_0=100$ and $g_1=80$.}
\end{figure}

\begin{figure}[tbp]
\includegraphics[width=8.8cm,]{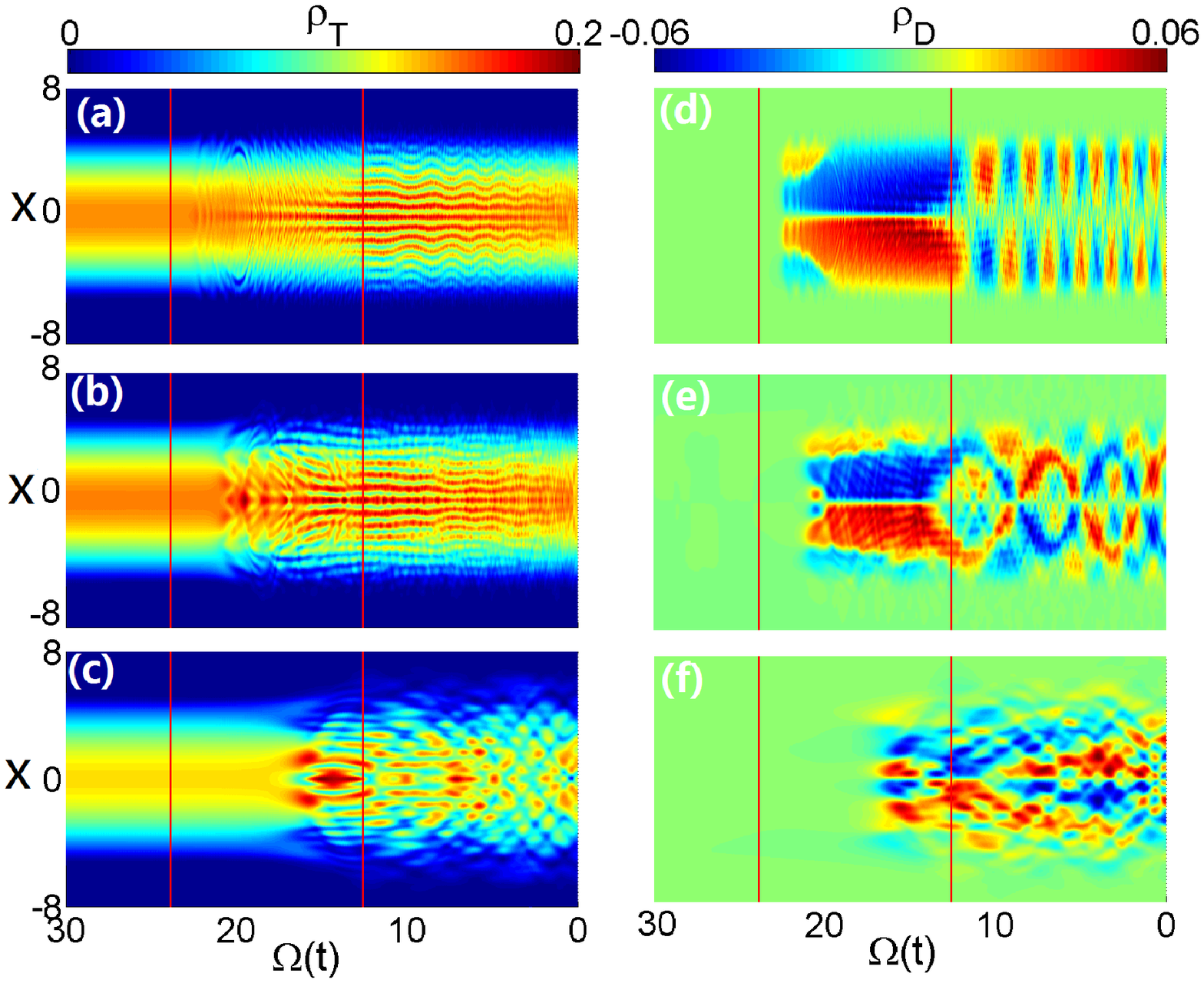}
\caption{(Color online) The density evolutions $\rho_T(x,t)$ and
$\rho_D(x,t)$ in the inverse sweep from phase III to phase II by
Raman driving $\Omega(t)=\Omega_0+\beta t$ with $\Omega_0=30$ and
typical sweep rates $\beta$: (a) and (d) $\beta=-0.1$, (b) and (e)
$\beta=-0.4$, (c) and (f) $\beta=-2$. The horizontal red lines
corresponds to the phase transition points. Other parameters are $k_0=5$, $g_0=100$ and $g_1=80$.}
\end{figure}

To further study the COM motion of the system, we calculate the
corresponding population evolution of the two spin components as
$P_{\sigma}(t)=\int_{-\infty}^{+\infty} |\psi_{\sigma}(x,t)|^2 dx $
and the COM evolution as
\begin{equation}
\langle x(t)\rangle=\int_{-\infty}^{+\infty}
\left[|\psi_{\uparrow}(x,t)|^{2}
+|\psi_{\downarrow}(x,t)|^{2}\right] xdx.
\end{equation}
As shown in Fig. 8(a,b), the atoms initially condense mainly in one
of two spins components, then their populations tend to be equal by
the Raman driving and finally exhibit periodical oscillation after
passing through the II-III phase transition point. However, both of
the period and the amplitude of the spin-population oscillation
decrease in time. As for the COM motion shown in Fig. 8(c,d), the
oscillation appears in the beginning even with very slow sweep [such
as $\beta=0.1$ in Fig. 8(c)], but its amplitude is very small
because the system almost follows its ground state. We note that the
oscillation in this case is similar to the well-known Zitterbewegung,
which has been observed in the SO-coupled BECs \cite{Qu,LeBlanc}.
After passing through the transition point, the oscillating COM motion recovers and the
oscillation amplitude is greatly enlarged and can be further
increased by increasing the Raman-driving rate.

At the end of this part, we consider the inverse sweep, i.e., the
sweep from phase III to phase II. The calculated density evolutions
$\rho_{T,D}(x,t)$ are shown in Fig. 9, where the initial state of
the system is in the phase III regime with $\Omega_0=30$ and the
Raman coupling is decreased linearly in time with the rates
$\beta=-0.1$ [Figs. 9(a,d)], $\beta=-0.4$ [Figs. 9(b,e)] and
$\beta=-2$ [Figs. 9(c,f)]. In these cases, the system can follow its
ground state with zero spin polarization at the beginning. As shown
in Fig. 9, the longitudinal spin polarization of the system becomes
finite and finally the excitations appear in the spin density with
periodic oscillation after the system passes through the III-II
phase boundary, and the excitations increase and the spin
oscillation becomes more complicated when speeding up the driving.
In this sweep, there is also no excited dark soliton for the same
reason.

\section{Conclusions}

In conclusion, we have numerically studied the ground states of
a 1D SO-coupled BEC in a harmonic trap and the Raman-driving
dynamics across different ground-state phases. We found the
coexistence of stripe and zero-momentum or plane wave phases in real
space due to the the competitions between repulsive interactions and
external trapping. We also showed a new method of dynamical
generation of dark solitons in the Raman-driving by sweeping the BEC through
the phase transition points. Both of the number and the stability of
the excited dark solitons can be controlled in this direct and
convenient way. Therefore, this system may provide a controllable
platform for creating dark solitons and studying their dynamics and
interaction properties. In view of the fact that the investigated
SO-coupled BEC with adjustable Raman coupling strength has been
realized by several experimental groups
\cite{Lin2011,Williams,Zhang2012,Qu,LeBlanc,Olson,Zhang2014}, it is
anticipated that our predictions in this work can be tested in an
experiment in the near future.

\begin{acknowledgments}
We thank Profs. Shi-Liang Zhu, Chaohong Lee and Biao Wu for many
helpful discussions. This work was supported by the NSF (Grant No. 11404108), the PCSIRT and the
SRFGS of SCNU. D. W. Z. acknowledges support from the postdoctoral
fellowship of HKU.
\end{acknowledgments}

\end{document}